\documentclass[seceq,preprint]{ptptex}

\newcommand{\bra}[1]{\left\langle\, #1\,\right|}
\newcommand{\ket}[1]{\left|\, #1\,\right\rangle}
\newcommand{\bracket}[2]{\left\langle\, #1\,|\,#2\,\right\rangle}
\newcommand{\VEV}[1]{\left\langle #1\right\rangle}
\newcommand{\wt}{\widetilde}
\newcommand{\wh}{\widehat}
\newcommand{\ol}{\overline}
\newcommand{\ra}{\rightarrow}

\newcommand{\nn}{\nonumber}
\newcommand{\para}[1]{\vspace*{1ex}
\noindent$\bullet~$\underline{\it #1}
\vspace*{.5ex}
}

\newcommand{\cH}{{\cal H}}
\newcommand{\cF}{{\cal F}}
\newcommand{\cK}{{\cal K}}
\newcommand{\cO}{{\cal O}}
\newcommand{\cT}{{\cal T}}

\newcommand{\bX}{{\bf X}}
\newcommand{\bP}{{\bf P}}
\newcommand{\bx}{{\bf x}}

\newcommand{\bp}{{\bf p}}
\newcommand{\bG}{{\bf\Gamma}}
\newcommand{\bM}{{\bf M}}
\newcommand{\id}{{\bf 1}}
\newcommand{\Dslash}{D\!\!\!\!\slash\,}

\newcommand{\del}{\partial}
\newcommand{\sq}{\sqrt{2}}

\newcommand{\AD}[1]{$\ol{\mbox{D~\,}}\!\!\!$#1}

\newcommand{\beq}{\begin{eqnarray}}
\newcommand{\eeq}{\end{eqnarray}}
\newcommand{\beqa}{\begin{eqnarray}}
\newcommand{\eeqa}{\end{eqnarray}}

\def\tr{\mathop{\rm tr}\nolimits}

\def\Tr{\mathop{\rm Tr}\nolimits}

\def\Str{\mathop{\rm Str}\nolimits}
\def\TrwhP{\mathop{\rm Tr\,\wh P}\nolimits}

\def\mat#1{\matt[#1]}
\def\matt[#1,#2,#3,#4]{\left(%
\begin{array}{cc} #1 & #2 \\ #3 & #4 \end{array} \right)}

\def\v2#1{\vv2[#1]}
\def\vv2[#1,#2]{\left(%
{#1 \atop #2}\right)}

\def\drawbox#1#2{\hrule height#2pt
        \hbox{\vrule width#2pt height#1pt \kern#1pt
              \vrule width#2pt}
              \hrule height#2pt}

\def\Fund#1#2{\vcenter{\vbox{\drawbox{#1}{#2}}}}
\def\Asymm#1#2{\vcenter{\vbox{\drawbox{#1}{#2}
              \kern-#2pt       
              \drawbox{#1}{#2}}}}

\def\fnd{\Fund{4.0}{0.4}}
\def\asym{\Asymm{4.0}{0.4}}
\def\sym{\fnd\kern-0.4pt\fnd}

\notypesetlogo  
\preprintnumber[3cm]{
hep-th/0305006\\
YITP-03-21\\
ITFA-2003-22\\
May, 2003
}

\title{
Exact description of D-branes in K-matrix theory
\footnote{Talk given by S.S. at the 17th Nishinomiya-Yukawa
Memorial Symposium on Nov.~23, 2002,
based on our recent work Ref.~\citen{AsSuTe3}.}
}

\author{
Tsuguhiko {\sc Asakawa},\footnote{
E-mail: asakawa@yukawa.kyoto-u.ac.jp}
Shigeki {\sc Sugimoto}\footnote{
E-mail: sugimoto@yukawa.kyoto-u.ac.jp}
and Seiji {\sc Terashima}\footnote{
E-mail: sterashi@science.uva.nl}
}

\inst{
${}^{**,\,***)}$
Yukawa Institute for Theoretical Physics,
Kyoto University,\\
Kyoto 606-8502, Japan\\
\vspace{2ex}
${}^{***)}$
The Niels Bohr Institute,\\
Blegdamsvej 17, DK-2100 Copenhagen \O, Denmark\\
\vspace{2ex}
${}^{\dag)}$
Institute for Theoretical Physics,
University of Amsterdam \\
Valckenierstraat 65,
1018 XE, Amsterdam, The Netherlands
}

\recdate{
}

\abst{
We summarize how to describe D-branes in a matrix theory based
on unstable D-instantons, which we call K-matrix theory,
and explicity show that D-branes can be constructed
as bound states of infinitly many unstable D-instantons.
We examine the fluctuations around D$p$-brane
solutions in the matrix theory and
show that they correctly reproduce fields
on the D$p$-brane world-volume.
Plugging them into the action of the matrix theory,
we precisely obtain the D$p$-brane action
as the effective action of the fluctuations.
}

\begin{document}

\maketitle

\section{Introduction}

In this paper, we consider matrix theories based on
unstable D-brane systems in type II string theory.
There are many choices of
the unstable D-brane systems. The lowest dimensional ones
are non-BPS D-instanton system in type IIA string theory
or D-instanton - anti D-instanton system in type IIB
string theory.
If one wants to avoid the formal Wick rotation
to obtain the D-instantons, one could start with
D$0$-\AD$0$ system in type IIA string theory or
non-BPS D$0$-brane system in type IIB string theory.
We call this kind of matrix theory as K-matrix theory.

There are many reasons we think K-matrix theory is interesting.
First of all, we can construct any D-brane configurations,
including BPS D-branes, non-BPS D-branes, D-brane
 - anti D-brane pairs, commutative as well as
 non-commutative D-branes, etc., as solutions
in the matrix theory.
This is in contrast to the usual approach in string theory,
in which D-branes are introduced by adding some open string
sectors by hand and one has to change the setup depending
on the brane configurations.
It may also be possible to construct these D-brane configurations
as solitonic solutions in open string field theory
associated with space-time filling unstable D-brane systems,
such as non-BPS D9-branes in type IIA
and D9-\AD9 system in type IIB or type I string theory.
But, K-matrix theory provides much simpler
formulation than these,
since the world-volume is just a point.

K-matrix theory is supposed to be a matrix formulation
of string theory such that creation
and annihilation of unstable D-branes are incorporated.
The creation and annihilation of D-branes is represented by
the value of tachyon, which is one of the matrix variables
in K-matrix theory. Roughly speaking,  D-branes are created
when the tachyon is around the top of its potential, 
while they are annihilated if the tachyon is at the bottom
of the potential. This tachyon is the very ingredient
which makes the theory quite different from other
matrix theories, such as Refs.~\citen{BFSS,IKKT},
and plays an essential role in the construction of D-branes.

We set the size of the matrix variables to be infinity
from the beginning. Thus, the matrix variables are considered
as operators acting on an infinite dimensional Hilbert space.
Therefore, the geometry of the D-brane world-volume is represented
in terms of operator algebras.
Remarkably, it turned out that each D-brane configuration
in K-matrix theory corresponds to an object
which is well-known among mathematicians.
As argued in Ref.~\citen{AsSuTe}, D-brane configurations are
given as (a limit of) spectral triples,
which are analytic description of Riemannian
manifold.\cite{Connes} \ 
Furthermore, the classification of D-brane
is naturally given by a group called analytic
K-homology. A physical interpretation of the
K-homology group in terms of K-matrix theory
has been given in Refs.~\citen{AsSuTe,AsSuTe2}.
These results are considered to be dual to 
the K-theory classification of D-brane charges
based on a physical interpretation
of the topological K-theory groups in terms of space-time
filling unstable D-brane.\cite{WittenK,Ho}

In this paper, we wish to explain more explicit description
of D-branes following our recent paper Ref.~\citen{AsSuTe3}.
We adopt the boundary string field theory action
of the unstable D-instantons as the action of K-matrix theory.
The exact solutions for this action representing D-branes
are given in Ref.~\citen{Te}. We will consider the fluctuations
around the D-brane solution and show that the effective
action of the fluctuations precisely agrees with the D-brane action.
The argument is also applicable to the construction of D-branes
as solitons in space-time filling unstable D-brane systems,
which enables us to generalize the results given in
Refs.~\citen{KuMaMo2,KrLa,TaTeUe}.
Our derivation is surprisingly simple and much more
powerful. It is powerful enough to show that not only the
D-brane tension, but also full effective action is
precisely reproduced without making any approximations.

\section{Boundary state formulation of BSFT action}
\label{bdryrev}

In this section, we quickly recapitulate some
of the ingredients in boundary string field theory (BSFT) action
for D-branes in type II superstring theory, which is
needed in the following discussion.
The BSFT action for the unstable D-brane system
in superstring was proposed in Refs.~\citen{KuMaMo2}
as a disk partition function with boundary interaction,
generalizing the usual definition
of the low energy effective action for massless fields
on a BPS D-brane, $i.e.$ Dirac-Born-Infeld action
including its derivative corrections.
\cite{AnTs,Wy} \
It can be expressed in terms of closed string variables as the
overlap between
the vacuum state and a boundary state with boundary interaction:
\footnote{See Ref.~\citen{AsSuTe3} for our convention and
the normalization.}
\begin{eqnarray}
S^{Dp}(A_\alpha,\Phi^i,T,\cdots)=\frac{2\pi}{g_s}
\bra{0} e^{-S_b(A_\alpha,\Phi^i,T,\cdots)} \ket{Bp;+}_{\rm NS}.
\label{BSFT}
\end{eqnarray}
Here $\ket{Bp;+}_{\rm NS}$ is the NS-NS sector boundary state
representing the D$p$-brane world-volume
and the open string fields $(A_\alpha,\Phi^i,T,\cdots)$
are turned on through the boundary interaction $S_b$.
Let us next explain these objects in the following.

The boundary state $\ket{Bp;\pm}_{\rm NS}$ used in (\ref{BSFT})
is defined by\cite{CLNY}
\begin{eqnarray}
\ket{Bp;\pm}_{\rm NS}=
\int[d\bx^\alpha]
\ket{\bx^\alpha, \bx^i=0;\pm}_{\rm NS},
\label{bpm}
\end{eqnarray}
where the superscript $\alpha=0,\dots,p$ and $i=p+1,\dots,9$
represent the directions
tangent and transverse to the D$p$-brane world-volume,
respectively.
Here we have introduced the boundary superfields
$\bx^\mu(\wh\sigma)=x^\mu(\sigma)+i\theta\,\psi^\mu(\sigma)$
~($\mu=0,\dots,9$),
where $x^\mu(\sigma)$ and $\psi^\mu(\sigma)$ are bosonic
and fermionic functions defined on the boundary of the disk
parametrized by $0\le\sigma\le 2\pi$ and
 $\wh\sigma=(\sigma,\theta)$ denotes the boundary
supercoordinate. The integral $\int[d\bx^\alpha]$ means
the path integral with respect to $x^\alpha(\sigma)$ and
$\psi^\alpha(\sigma)$.
The state $\ket{\bx^\mu;\pm}_{\rm NS}$ is the coherent state
satisfying
\begin{eqnarray}
\bX^\mu(\wh\sigma)\ket{\bx^\mu;\pm}_{\rm NS}=
\bx^\mu(\wh\sigma)\ket{\bx^\mu;\pm}_{\rm NS},
\label{coherent}
\end{eqnarray}
where $\bX^\mu(\wh\sigma)=X^\mu(\sigma)+i\theta\,\Psi^\mu_\pm(\sigma)$
is a superfield operator whose components
$X^\mu(\sigma)=\wh x_0^\mu+i\sum_{m\ne 0}
(\frac{\alpha_m^\mu}{m}e^{-im\sigma}+\frac{\tilde\alpha_m^\mu}{m}
e^{im\sigma})$ and
$\Psi^\mu_\pm(\sigma)=\sum_{r}(\Psi^\mu_r e^{-ir\sigma}
\pm i\wt\Psi^\mu_r e^{ir\sigma})$
are closed string operators in NS-R formulation
evaluated at the boundary of the disk.
We use NS-NS sector boundary interaction in (\ref{BSFT})
and hence the subscript $r$ of $\Psi^\mu_r$ and $\wt\Psi^\mu_r$
runs half odd integers. We often omit the subscript $\pm$ of
the fermion $\Psi_\pm^\mu$ in the following.
The path integral in the right hand side of (\ref{bpm}) represents
summing over all possible configurations of the boundary of the 
string world-sheet attached on the D$p$-brane world-volume.

It is also useful to introduce a momentum superfield operator
$\bP_\mu(\wh\sigma)=\theta\,P_\mu(\sigma)+i\Pi_\mu(\sigma)$,
where $P_\mu(\sigma)$ and $\Pi_\mu(\sigma)$ are the conjugate
momenta of $X^\mu(\sigma)$ and $\Psi^\mu(\sigma)$, respectively,
and write the coherent state as
\begin{eqnarray}
\ket{\bx^\mu;\pm}_{\rm NS}=e^{-i\int\!
 d\hat\sigma\, \bP_\mu\bx^\mu}\ket{\bx^\mu=0;\pm}_{\rm NS},
\label{shift}
\end{eqnarray}
where $d\wh\sigma=d\sigma d\theta$.
{}From this expression, we can see that
the boundary state (\ref{bpm}) for a D$p$-brane
is related to that for a D$(-1)$-brane
 $\ket{B(-1);\pm}_{\rm NS}=\ket{\bx^\mu=0;\pm}_{\rm NS}$ by
\begin{eqnarray}
\ket{Bp;\pm}_{\rm NS}=
\int[d\bx^\alpha]\,e^{-i\int\!
 d\hat\sigma\, \bP_\alpha\bx^\alpha}\ket{B(-1);\pm}_{\rm NS}.
\label{bpm2}
\end{eqnarray}

The fields on the D-brane couple to the boundary
of the string world-sheet,
and the disk partition function
changes its value when the fields on the D-brane are
turned on. The coupling is represented by the boundary interaction
$S_b$ in (\ref{BSFT}).
The boundary interaction for the gauge field $A_\alpha(x)$ on the
D-brane is well-known\cite{CLNY,AnTs} and given as
a supersymmetric generalization of Wilson loop operator:
\begin{eqnarray}
e^{-S_b(A_\alpha)}
=\TrwhP \exp\left(-\int
d\wh\sigma \,A_\alpha(\bX)D\bX^\alpha
\right),
\label{Sgauge}
\end{eqnarray}
where the covariant derivative is defined as
$D=\del_\theta+\theta\del_\sigma$.
Here $\wh{\rm P}$ denotes the supersymmetric generalization of
the path ordered product, which is defined as
\begin{eqnarray}
&&\wh{\rm P}\,\exp\left(\int d\wh\sigma \,\bM(\wh\sigma)\right)\nn\\
&=&
\sum_{n=0}^\infty(-1)^{\frac{n(n-1)}{2}}
\int d\wh\sigma_1\cdots d\wh\sigma_n
\Theta(\wh\sigma_{12})\Theta(\wh\sigma_{23})\cdots
\Theta(\wh\sigma_{n-1\,n})
\,\bM(\wh\sigma_1)\cdots\bM(\wh\sigma_n),\nn\\
\label{SP}
\end{eqnarray}
where $\wh\sigma_{ab}=\sigma_a-\sigma_b-\theta_a\theta_b$
and $\Theta$ is a step function.
When the gauge field is abelian and the field strength
$F_{\alpha\beta}$ is constant, we can explicitly calculate
the expression (\ref{BSFT}) with the boundary interaction
(\ref{Sgauge}) and obtain the DBI action.

On the D$p$-brane with $p<9$, we also have massless
scalar fields $\Phi^i(x)$ which represent the transverse position of
the D$p$-brane world-volume. Since the vertex operators
correspond to the scalar fields are given by the momentum
operator $\bP_i$, the boundary interaction including the
scalar fields will become
\begin{eqnarray}
e^{-S_b(A_\alpha,\Phi^i)}
=\TrwhP \exp\left\{-\int
d\wh\sigma 
\left(A_\alpha(\bX)D\bX^\alpha
+i\Phi^i(\bX)\bP_i\right)
\right\}.
\label{BPSbdry}
\end{eqnarray}
In fact, we can see from (\ref{shift}) that turning on
the scalar fields $\Phi^i$ in the boundary interaction
(\ref{BPSbdry}) corresponds to shifting
the D-brane world-volume in the transverse directions.
In this paper, we will not turn on fermion fields and
massive fields on the D$p$-brane.
Thus, (\ref{BPSbdry}) contains all what we need to
describe BPS D$p$-branes.

It is not difficult to include tachyon fields
to obtain the boundary interaction corresponding
to D$p$-\AD$p$ systems or non-BPS
D$p$-branes.\cite{HaKuMa,KuMaMo2,KrLa,TaTeUe} \ 
First, we introduce a matrix consists of a condensate of
the open string vertices in the superfield notation
\begin{eqnarray}
\bM=\mat{-A_\alpha(\bX) D\bX^\alpha
-i\Phi^i(\bX)\bP_i,T(\bX),T(\bX)^\dag,
-\wt A_\alpha(\bX) D\bX^\alpha-i\wt\Phi^i(\bX)\bP_i}.
\label{superM}
\end{eqnarray}
Here $A_\alpha$, $\wt A_\alpha$, $\Phi^i$, $\wt\Phi^i$ and $T$
are the fields on the D-brane.
For D$p$-\AD{$p$} system,
these fields are independent, and
$A_\alpha$ and $\Phi^i$ represent the gauge field and scalar fields
on the D$p$-brane,
$\wt A_\mu$ and $\wt\Phi^i$ are those on the
\AD{$p$}-brane and $T$ is the tachyon field which
is created by the open string stretched between
the D$p$-brane and the \AD{$p$}-brane.
For non-BPS D$p$-brane, we have constraints
 $A_\alpha=\wt A_\alpha$,
$\Phi^i=\wt\Phi^i$ and $T^\dag=T$.

Then the boundary interaction (for NS-NS sector) is given by
\begin{eqnarray}
e^{-S_b}&=&
\kappa\TrwhP\,e^{\int\! d\hat\sigma\, \bM(\hat\sigma)}
\label{supertrP}
\end{eqnarray}
where $\bM$ is given as (\ref{superM}).
The normalization constant $\kappa$ is 
$\kappa=1$ for D$p$-\AD$p$ systems and $\kappa=1/\sqrt{2}$
for non-BPS D-branes.

It is often useful
to rewrite the boundary interaction
(\ref{supertrP}) in path integral formulation.
Let us suppose that the number of the D$p$-branes and
\AD$p$-branes are both $2^{m-1}$ and the matrix $\bM$ in (\ref{superM})
can be expanded by $SO(2m)$ gamma matrices
$\Gamma^I=\left({~~~\gamma^I\atop\gamma^{I\dag}~~~}\right)$
~($I=1,\dots,2m$) as
\begin{eqnarray}
\bM=\sum_{k=0}^{2m} \bM^{I_1\cdots I_k}\,
\Gamma^{I_1\cdots I_k},
\label{gammaexp}
\end{eqnarray}
where
 $\Gamma^{I_1\cdots I_k}$ denote the
skew-symmetric products of the gamma matrices
and $\bM^{I_1\cdots I_k}$ are the coefficients.
In this case, the boundary interaction becomes\cite{KrLa}
\begin{eqnarray}
e^{-S_b}=\int[d\bG^I]\,
\exp\left\{
\int d\wh\sigma \left(
\frac{1}{4}\bG^I D\bG^I
+\sum_{k=0}^{2m} \bM^{I_1\cdots I_k}
\bG^{I_1}\cdots\bG^{I_k}
\right)\right\}.
\label{super}
\end{eqnarray}
where
 $\bG^I(\wh\sigma)=\eta^I(\sigma)+\theta\,F^I(\sigma)$
are real fermionic superfields. Note that we impose anti-periodic
boundary condition for $\eta^I(\sigma)$ for the
NS-NS sector boundary interaction.
This formula is obtained from (\ref{supertrP})
by replacing the gamma
matrices $\Gamma^I$ in the gamma matrix expansion (\ref{gammaexp})
with the superfields $\bG^I$
and arranging the kinetic term $\frac{1}{4}\bG^ID\bG^I$ for them.
For example, for the case with a single non-BPS D$p$-brane,
the matrix $\bM$ in (\ref{superM}) is expanded by $\sigma^1$
and the boundary interaction (\ref{super}) will
become\cite{HaKuMa,KuMaMo2}
\begin{eqnarray}
&&e^{-S_b(A_\alpha,\Phi^i,T)}\nn\\
&=&\int[d\bG]\,
\exp\left\{
\int d\wh\sigma \left(
\frac{1}{4}\bG D\bG
-A_\alpha(\bX)D\bX^\alpha-i\Phi^i(\bX)\bP_i+
T(\bX)\bG\right)\right\}.
\label{nonBPSbdry}
\end{eqnarray}

The equivalence of (\ref{supertrP}) and (\ref{super})
follows from the equivalence between
the operator formulation and the path integral formulation
of boundary supersymmetric quantum
mechanics.\footnote{Note however that the supersymmetry is
explicitly broken by the
anti-periodic boundary condition of the boundary fermions
in the NS-NS sector.} \ 
See appendix A of Ref.~\citen{AsSuTe3} for a formal proof.
The essential point is as follows. The kinetic term
of $\bG^I$ in (\ref{super}) implies the canonical
anti-commutation relation 
\begin{eqnarray}
\{\eta^I,\eta^J\}=2\delta^{IJ}
\label{car}
\end{eqnarray}
in the operator formulation, and hence 
$\eta^I$ play the same role as the gamma matrices $\Gamma^I$.
In order to obtain a supersymmetric formula,
it is natural to combine the boundary fermion $\eta^I(\sigma)$
with the bosonic auxiliary field $F^I(\sigma)$ into the
superfield $\bG^I(\wh\sigma)$.
This is the reason that we replace $\Gamma^I$ with $\bG^I$
when we rewrite (\ref{supertrP}) to (\ref{super}).

\section{D-brane configurations in K-matrix theory}
\label{DinKmat}
In this section, we consider the construction of D-branes
in K-matrix theory. We will use the matrix theory based
on the non-BPS D-instantons in type IIA string theory
for simplicity.
The generalization to other unstable D-brane systems
is straightforward.

\subsection{D-brane solutions in K-matrix theory}

Let us consider
 the matrix theory based on
non-BPS D-instantons in type IIA string theory.
The field contents and the action is obtained as
the dimensional reduction of non-BPS D9-brane system.
The system of $N$ non-BPS D9-branes is a ten dimensional
$U(N)$ gauge theory with a tachyon field.
In this paper, we only consider tachyonic and massless
bosonic fields, that is, the tachyon field $T$
and the gauge field $A_\mu$, both of which transform
as the adjoint representation under the gauge transformation.
The dimensional reduction of these fields gives
the tachyon $T$ and ten scalars $\Phi^\mu$ ($\mu=0,1,\dots,9$),
which are $N\times N$ hermitian matrix variables
in the matrix theory.
We take $N$ to be infinity
and regard these matrices as operators acting on an infinite
dimensional Hilbert space.
For the action of this system,
we adopt the BSFT action (\ref{BSFT})
\begin{eqnarray}
S^{D(-1)}(\Phi^\mu,T)=\frac{2\pi}{g_s}\bra{0}e^{-S_b(\Phi^\mu,T)}
\ket{B(-1);+}_{\rm NS},
\label{Kmatrixaction}
\end{eqnarray}
where the boundary interaction is now given by (\ref{supertrP})
with
\begin{eqnarray}
\bM=\mat{-i\Phi^\mu\bP_\mu,T,
T,-i\Phi^\mu\bP_\mu}.
\label{bM}
\end{eqnarray}

A configuration representing
a D$p$-brane extended along $x^0,\dots,x^p$-directions
is given by
\begin{eqnarray}
T&=&u\,\sum_{\alpha=0}^p
\wh{p}_\alpha\, \gamma^\alpha
\label{sol1-1}
\\
\Phi^{\alpha} &=& \wh{x}^{\alpha}
\quad
(\alpha=0, \dots, p)\,,
\qquad
\Phi^i=0
\quad
(i=p+1, \dots, 9)\,,
\label{sol1-2}
\end{eqnarray}
where $\wh x^\alpha$ and $\wh p_\alpha$ are operators
on a Hilbert space $\cH$ satisfying
\begin{eqnarray}
[\wh x^\alpha,\wh p_\beta]=i\delta^\alpha_\beta,
\label{ccr}
\end{eqnarray}
and $\gamma^\alpha$ are hermitian gamma matrices.
$T$ and $\Phi^\mu$ are operators acting on a Hilbert space
$\cH\otimes S$, where $S$ is the spinor space on which
the gamma matrices $\gamma^\alpha$ are represented.
$u$ is a real parameter and this configuration becomes
an exact solution in the limit $u\ra \infty$.\cite{Te}

It will soon become clear in section \ref{action}
that this configuration becomes an exact solution
representing a D$p$-brane (BPS D$p$-brane or non-BPS D$p$-brane
for even or odd $p$, respectively).
Before moving to the discussion using the action,
let us explain some of the properties that we can immediately
see from the configuration (\ref{sol1-1},\ref{sol1-2}).

Recall that the eigen values of $\Phi^\mu$ represent the
position of the non-BPS D-instantons.
Since the spectrum of the operator $\wh x^\alpha$
spans the real axis, we can see from (\ref{sol1-2})
that this configuration represent a $p+1$ dimensional
object, which is interpreted as the D$p$-brane world-volume.
In particular, the eigen value $x^\alpha$
of the operator $\wh x^\alpha$ is interpreted as a coordinate
on the D$p$-brane world-volume.

This matrix theory has a huge gauge symmetry:
\begin{eqnarray}
\Phi^\mu\ra U\Phi^\mu U^{-1},~~~~T\ra U T U^{-1},
\label{gaugesym}
\end{eqnarray}
where $U$ is a unitary transformation of the Chan-Paton
Hilbert space on which the operators $\Phi^\mu$ and $T$ are defined.
The configuration (\ref{sol1-1},\ref{sol1-2}) breaks
most of the symmetry, but the unitary transformations
of the form
\begin{eqnarray}
U=h\,\id,
\label{U1gauge}
\end{eqnarray}
where $\id$ is the identity operator on the Hilbert space
and $h$ is the $U(1)$ phase factor,
are left unbroken. This unbroken $U(1)$ subgroup is interpreted
as the global gauge symmetry of the D$p$-brane.
If we repeat the same argument for 
a configuration representing $N$ D$p$-branes,
which can be obtained
by piling $N$ copies of a single D$p$-brane configuration
(\ref{sol1-1},\ref{sol1-2}), we obtain $U(N)$ as the unbroken
subgroup. This is consistent with
the global gauge symmetry of $N$ D$p$-branes.

\subsection{Fluctuations around the D-brane solution}
\label{general}

Let us next consider the fluctuations around the D$p$-brane solution
(\ref{sol1-1},\ref{sol1-2}).
Here we mainly consider the cases with even $p$,
in which (\ref{sol1-1},\ref{sol1-2}) represents a BPS D$p$-brane,
for simplicity.
General fluctuations will be given by
\begin{eqnarray}
T&=&u\,\sum_{\alpha=0}^p
\wh{p}_\alpha\, \gamma^\alpha+\delta T,
\label{fluc1-1}
\\
\Phi^{\alpha} &=& \wh{x}^{\alpha}+\delta \Phi^\alpha
\quad\quad
(\alpha=0, \dots, p)\,,\\
\Phi^i&=&\delta \Phi^i
\quad\quad\quad\quad~
(i=p+1, \dots, 9)\,.
\label{fluc1-2}
\end{eqnarray}
Fields on the D$p$-brane are embedded in
the fluctuations $\delta T$ and $\delta\Phi^\mu$.
In principle, we have to deal with all the possible fluctuations,
but there is a shortcut to extract the components corresponding to
massless fields on the D$p$-brane.
The key observation is that the massless fluctuations can be
interpreted as Nambu-Goldston modes associated with the
symmetry of the system.

For example, the matrix theory has a translational symmetry:
\begin{eqnarray}
\Phi^i\ra\Phi^i+\phi^i\,\id,
\label{transl}
\end{eqnarray}
where $\phi^i$ is a real number. This symmetry is
spontaneously broken in the configuration (\ref{sol1-2}).
Then, the NG mode $\delta\Phi^i=\phi^i(\wh x)$
corresponds to a massless scalar field on the D$p$-brane.
Here we allow $\wh x^\alpha$ dependence for $\phi^i$,
since we know the eigen value of $\wh x^\alpha$
represent a coordinate of the D$p$-brane world-volume.
Note that similar shift (\ref{transl}) for $\Phi^\alpha$ can be
absorbed by the gauge transformation (\ref{gaugesym}) and
the corresponding fluctuation
$\delta\Phi^\alpha=\phi^\alpha(\wh x)$ does not represent
a physical field.
In fact, the gauge transformation (\ref{gaugesym}) with
$U=\exp(i\{\wh p_\alpha,\lambda^\alpha(\wh x)\}/2)$ implies
\begin{eqnarray}
U\wh x^\alpha U^{-1}=\wh x^\alpha+\lambda^\alpha(\wh x)
+{\cal O}(\lambda^2)
\end{eqnarray}
and we can always set $\phi^\alpha=0$ by choosing
$\lambda^\alpha$ appropriately.
This represents the reparametrization invariance of the D$p$-brane
world-volume. Setting $\phi^\alpha=0$ corresponds
to choosing the static gauge on the D$p$-brane.
It is amusing that both the reparametrization of
the world-volume coordinates and the gauge transformation
of the D$p$-brane are originated in the gauge transformation
(\ref{gaugesym}) of the matrix theory.

Next, let us consider the fluctuation corresponding to
the gauge field on the D$p$-brane. The gauge field is
associated with the local gauge symmetry on the D$p$-brane
world-volume. The local gauge symmetry is obtained by
allowing $\wh x^\alpha$ dependence for the $U(1)$ factor $h$
in (\ref{U1gauge}). This symmetry is spontaneously broken,
as we can see that the tachyon operator in (\ref{sol1-1})
doesn't commute with $h(\wh x)$:
\footnote{
Of course, this doesn't mean the theory is in Higgs phase,
since the global gauge symmetry with $h(x)={\rm constant}$
is unbroken. Note also that our argument here is limited to
classical theory.}
\begin{eqnarray}
h(\wh x)\,T\,h(\wh x)^{-1}=T-iu\, h(\wh x)\del_\alpha h(\wh x)^{-1}
\gamma^\alpha.
\label{inhomo}
\end{eqnarray}
The gauge field can be considered as a NG mode for
the spontaneously broken `local' gauge symmetry with
$h(x)\del_\alpha h(x)^{-1}={\rm constant}$, and hence
it corresponds to the fluctuation
of the form $\delta T=-iu A_\alpha(\wh x)\gamma^\alpha$.
It is now clear that the second term in the right hand side
of (\ref{inhomo}) can be absorbed in the fluctuation
$A_\alpha(\wh x)$ provided it transforms as
\begin{eqnarray}
A_\alpha(\wh x)\ra A_\alpha(\wh x)
+ h(\wh x)\del_\alpha h(\wh x)^{-1}
\end{eqnarray}
under the gauge transformation, which guarantees the local gauge
invariance of the D$p$-brane world-volume theory.

Including these fluctuations, the D$p$-brane configuration becomes
\begin{eqnarray}
T&=&u\,\sum_{\alpha=0}^p
\left(\wh{p}_\alpha-i A_\alpha(\wh x)\right)\,
\gamma^\alpha\,,
\label{fluc2-1}
\\
\Phi^{\alpha} &=& \wh{x}^{\alpha}
\quad\quad
(\alpha=0, \dots, p)\,,
\label{fluc2-2}\\
\Phi^i&=&\phi^i(\wh x)
~~~
(i=p+1, \dots, 9)\,.
\label{fluc2-3}
\end{eqnarray}
This is the configuration representing a BPS D$p$-brane
with massless bosonic fields turned on.

When $p$ is odd, the solution
(\ref{sol1-1},\ref{sol1-2}) represents a non-BPS D$p$-brane,
and hence, we expect a fluctuation corresponding to the tachyon
field on it. Actually, we can find the tachyonic fluctuation
in the tachyon operator as
\begin{eqnarray}
T&=&u\,\sum_{\alpha=0}^p
\left(\wh{p}_\alpha-i A_\alpha(\wh x)\right)\,
\gamma^\alpha+ t(\wh x)\,\gamma_*,
\label{tachyonfluc}
\end{eqnarray}
where $t(x)$ is a real scalar function and
$\gamma_*$ is the chirality operator satisfying
\begin{eqnarray}
\gamma_*^2=1,~~~~
\{\gamma^\alpha,\gamma_*\}=0~~~
(\alpha=0,\dots,p).
\label{gamma}
\end{eqnarray}
Note that such $\gamma_*$ exists only when $p$ is odd.
The fluctuation $t(\wh x)$ in (\ref{tachyonfluc})
corresponds to the tachyon field on the non-BPS D$p$-brane.
To understand this quickly, consider
the case with $A_\alpha(x)=0$ and $t(x)=t$ (constant),
for simplicity. Then, using (\ref{gamma}), we obtain
\begin{eqnarray}
T^2=u^2\wh p_\alpha^2+ t^2,
\end{eqnarray}
from which it immediately follows that
the field $t$ has a negative mass squared mass term
in the effective action, which is inherited from that for $T$.
This can be more explicitly seen by plugging the 
configuration (\ref{tachyonfluc})
into the action (\ref{Kmatrixaction}) as we will explain
in the next section. 

One might be wondering whether or not
more (unwanted) fields are introduced when we turn on
general fluctuations (\ref{fluc1-1})--(\ref{fluc1-2}).
As explained in Ref.~\citen{AsSuTe3}, the
fluctuations introduced in (\ref{fluc2-1})--(\ref{fluc2-3})
and (\ref{tachyonfluc}) are essentially the only relevant
ones in the limit $u\ra\infty$.
The others will disappear in the effective action
when we take the limit $u\ra\infty$.
We will come back to this point in the next section.

\section{D$p$-brane action from K-matrix Theory}
\label{action}
\subsection{Derivation of D$p$-brane action}
\label{ascent}
The effective action of the fluctuations around the D$p$-brane
solution is obtained by plugging the configuration
(\ref{fluc2-1})--(\ref{fluc2-3})
or (\ref{tachyonfluc}) into the action (\ref{Kmatrixaction})
of K-matrix theory.
Here we will demonstrate this
for the configuration (\ref{fluc2-1})--(\ref{fluc2-3})
representing a BPS D$p$-brane.
Inserting (\ref{fluc2-1})--(\ref{fluc2-3})
into (\ref{bM}), we obtain
\begin{eqnarray}
\bM=-i\wh x^\alpha\bP_\alpha-i\phi^i(\wh x)\bP_i+
 u(\wh p_\alpha-iA_\alpha(\wh x))\Gamma^\alpha,
\label{bMDp}
\end{eqnarray}
where $\Gamma^\alpha=\left({~~~\gamma^\alpha\atop\gamma^\alpha~~~}\right)$.
The boundary interaction is given as (\ref{supertrP})
 with this $\bM$ in the operator formulation.
The key step is to rewrite this boundary interaction
in path integral formulation. Using the same argument to obtain
(\ref{super}), the boundary interaction is now rewritten as
\begin{eqnarray}
e^{-S_b(\Phi^\mu,T)}&=&\int[d\bG^\alpha][d\bx^\alpha][d\bp_\alpha]
\exp\left\{
\int d\wh\sigma \left(
\frac{1}{4}\bG^\alpha D\bG^\alpha+i\bp_\alpha D\bx^\alpha
\right.\right.\nn\\
&&\left.\left.
~~-i\bx^\alpha\bP_\alpha-i\phi^i(\bx)\bP_i
+u(\bp_\alpha-iA_\alpha(\bx))\bG^\alpha
\frac{}{}\right)\right\},
\label{pathint}
\end{eqnarray}
where 
$\bx^\alpha(\wh\sigma)=x^\alpha(\sigma)+i\theta\psi^\alpha(\sigma)$
and $\bp_\alpha(\wh\sigma)=p_\alpha(\sigma)+i\theta\pi_\alpha(\sigma)$
are boundary superfields.
To obtain (\ref{pathint}),
we replaced the operators $\wh x^\alpha$,
$\wh p_\alpha$ and $\Gamma^\alpha$ with the superfields
$\bx^\alpha$, $\bp_\alpha$ and $\bG^\alpha$, respectively,
and arranged the kinetic terms $\frac{1}{4}\bG^\alpha D\bG^\alpha$
and $i\bp_\alpha D\bx^\alpha$ to realize the (anti-) commutation
relations (\ref{car}) and (\ref{ccr}), respectively.
The equivalence of (\ref{supertrP}) and (\ref{pathint})
again follows from the equivalence between
the operator formulation and the path integral formulation
of boundary supersymmetric quantum mechanics.
The formal derivation of this can be found in
the appendix A of Ref.~\citen{AsSuTe3}.

Now we are ready to derive the D$p$-brane action.
It is surprisingly simple.
First, note that we can
perform the path integral over $\bp_\alpha$,
which implies a delta functional imposing the condition
\begin{eqnarray}
iD\bx^\alpha+u\bG^\alpha=0.
\label{Dx}
\end{eqnarray}
Integrating with respect to $\bG^\alpha$, we obtain
\begin{eqnarray}
&&e^{-S_b(\Phi^\mu,T)}\nn\\
&=&\int[d\bx^\alpha]
\exp\left\{-\int d\wh\sigma \left(
\frac{1}{4u^2}D\bx^\alpha D^2 \bx^\alpha
+i\bx^\alpha\bP_\alpha+i\phi^i(\bx)\bP_i
+A_\alpha(\bx)D\bx^\alpha\right)\right\}.
\nn\\
\label{superbdry}
\end{eqnarray}
The first term $\frac{1}{4u^2}D\bx^\alpha D^2\bx^\alpha$
in the integrand of (\ref{superbdry}) drops in the limit
$u\ra\infty$. Then, using (\ref{bpm})--(\ref{bpm2}), we have
\begin{eqnarray}
e^{-S_b(\Phi^\mu,T)}\ket{B(-1);\pm}_{\rm NS}
&\ra& e^{-S_b(A_\alpha,\phi^i)}
\ket{Bp;\pm}_{\rm NS},~~~~(u\ra\infty),
\label{Dp2}
\end{eqnarray}
where
\begin{eqnarray}
e^{-S_b(A_\alpha,\phi^i)}
&=&\exp\left\{-\int d\wh\sigma \left(
i\phi^i(\bX)\bP_i+A_\alpha(\bX)D\bX^\alpha
\right)\right\}.
\label{Dp3}
\end{eqnarray}
The boundary interaction (\ref{Dp3}) is nothing but
that for a D$p$-brane given in (\ref{BPSbdry})!
In particular, if we apply this result to the action
of the form (\ref{Kmatrixaction}), we precisely obtain
the BSFT action for the D$p$-brane (\ref{BSFT}):
\begin{eqnarray}
S^{D(-1)}(\Phi^\mu,T)\ra S^{Dp}(A_\alpha,\phi^i),
~~~~(u\ra\infty).
\end{eqnarray}
Thus, we correctly reproduced the D$p$-brane action from
K-matrix theory.

\subsection{Some comments}

Here we make several comments on the derivation of D$p$-brane
action given in the previous subsection.

\para{On the $u\ra\infty$ limit}

To be more precise, we should regularize the path integral
before taking $u\ra\infty$ limit. If we keep $u$ finite, we obtain
\begin{eqnarray}
&&e^{-S_b(\Phi^\mu,T)}\ket{B(-1);\pm}_{\rm NS}\nn\\
&=&e^{-S_b(A_\alpha,\phi^i)}
\int[d\bx^\alpha]\,
e^{-\int d\hat\sigma\,\frac{1}{4u^2}D\bx^\alpha D^2\bx^\alpha}
\ket{\bx^\alpha,\bx^i=0;+}_{\rm NS}.
\end{eqnarray}
Here the path integral with respect to $\bx^\alpha$
only involves gaussian integral and it can be performed
by using zeta-function regularization. The result
justifies the naive $u\ra\infty$ limit
we took in (\ref{Dp2}).\cite{AsSuTe} \ 
If we turn off the fluctuations $A_\alpha$ and $\phi^i$,
the gaussian path integral with zeta function regularization implies
\begin{eqnarray}
S^{D(-1)}(\Phi^\mu,T)
&=&\frac{2\pi}{g_s}\int[d\bx^\alpha]\,
e^{-\int d\hat\sigma\,\frac{1}{4u^2}D\bx^\alpha D^2\bx^\alpha}
\VEV{0\,\big|\,\bx^\alpha,\bx^i=0;+}_{\rm NS}\nn\\
&=&
\cK(u^2)^{p+1}\,\cT_p\,V\,,
\end{eqnarray}
where $V\equiv\int\!d^{p+1}x$ is the volume of the D$p$-brane
world-volume,
$\cT_p\equiv
S^{Dp}|_{A_\alpha=\phi^i=0}/V$ is the D$p$-brane tension
and
\begin{eqnarray}
\cK(z)=
\frac{\sqrt{z}\,4^{z}\Gamma(z)^2}{2\sqrt{\pi}\Gamma(2z)}\,.
\end{eqnarray}
This is the function which appeared in Ref.~\citen{KuMaMo2},
in which it was shown that the tension of D-branes
constructed via tachyon condensation
in supersymmetric BSFT
precisely agrees with the correct value.
In their argument,
the fact that $\cK(u^2)$ becomes $1$ in the $u\ra\infty$ limit
was crucial to obtain the correct tension.
Our argument provides an easy way to understand this fact.
Once we accept the naive $u\ra\infty$ limit (\ref{Dp2}),
there is nothing mysterious and we can directly come to the same
conclusion without performing detailed calculation.

\para{Generalization to non-BPS D-branes}

It is straightforward to generalize the above argument
to include the tachyon fluctuation as (\ref{tachyonfluc})
for the $p={\rm odd}$ case with a non-BPS D$p$-brane.
When we insert (\ref{tachyonfluc}) in (\ref{bM}), (\ref{bMDp})
is replaced with
\begin{eqnarray}
\bM=-i\wh x^\alpha\bP_\alpha-i\phi^i(\wh x)\bP_i+
 u(\wh p_\alpha-iA_\alpha(\wh x))\Gamma^\alpha+t(\wh x)\Gamma_*,
\end{eqnarray}
where $\Gamma_*=\left({~~~\gamma_*\atop\gamma_*~~~}\right)$,
and the boundary interaction becomes
\begin{eqnarray}
e^{-S_b(\Phi^\mu,T)}&=&\int[d\bG^I][d\bx^\alpha][d\bp_\alpha]
\exp\left\{
\int d\wh\sigma \left(
\frac{1}{4}\bG^I D\bG^I+i\bp_\alpha D\bx^\alpha
\right.\right.\nn\\
&&\left.\left.
~~-i\bx^\alpha\bP_\alpha-i\phi^i(\bx)\bP_i
+u(\bp_\alpha-iA_\alpha(\bx))\bG^\alpha+t(\bx)\bG_*
\frac{}{}\right)\right\},
\end{eqnarray}
where $\bG^I=(\bG^\alpha,\bG_*)$. The same argument as
in section \ref{ascent} implies
\begin{eqnarray}
e^{-S_b(\Phi^\mu,T)}\ket{B(-1);\pm}_{\rm NS}
&\ra& e^{-S_b(A_\alpha,\phi^i,t)}
\ket{Bp;\pm}_{\rm NS},~~~~(u\ra\infty)
\label{nonDp2}
\end{eqnarray}
with
\begin{eqnarray}
&&e^{-S_b(A_\alpha,\phi^i,t)}\nn\\
&=&\int[d\bG_*]\,\exp\left\{
\int d\wh\sigma \left(
\frac{1}{4}\bG_* D\bG_*
-i\phi^i(\bX)\bP_i
-A_\alpha(\bX)D\bX^\alpha+t(\bX)\bG_*
\right)\right\}.\nn\\
\label{nonDp3}
\end{eqnarray}
As expected, (\ref{nonDp3}) is nothing but the boundary interaction
for a non-BPS D-brane given in (\ref{nonBPSbdry}).

It is also easy to generalize this argument to
the construction of multiple D-branes,
D-brane - anti D-brane systems, non-commutative D-branes
and so on.

\para{More general fluctuations}

In (\ref{bMDp}), we only considered the fluctuations up to
linear order with respect to the gamma matrices.
We could turn on more general fluctuations that induce
higher order terms in the gamma matrix expansion (\ref{gammaexp}).
However, since (\ref{Dx}) implies that $\bG^\alpha$ is a variable
of order $\cO(u^{-1})$, the higher order terms will vanish
in the $u\ra\infty$ limit.
\footnote{Here we assumed the explicit $u$ dependence
of the fluctuations is
$\delta T\sim\cO(u^1)$ and $\delta\Phi\sim\cO(u^0)$
so that the fluctuations
do not grow faster than the D$p$-brane solution itself
in the $u\ra\infty$ limit.}

\para{Coupling to closed string fields}

In (\ref{Dp2}) or (\ref{nonDp2}) 
we reproduced the D$p$-brane boundary states
with boundary interaction. The boundary states carry
much more information than the effective action
(\ref{BSFT}).
In fact, coupling of the D$p$-brane to closed string field
$\varphi$
can be expressed as $\bracket{\varphi}{Bp}$.
Therefore, our argument in the previous subsection
shows that not only the effective action of the D$p$-brane,
but also its coupling to every closed string field
is precisely reproduced from the configuration
(\ref{fluc2-1})--(\ref{tachyonfluc}) in K-matrix theory.
One of the interesting applications of this fact will 
be given in the next section.

\para{Application to the D-brane descent relations}

Our method can also be applied to the construction of D-branes
via tachyon condensation in the higher dimensional
unstable D-brane systems.
For example, a BPS D$p$-brane configuration in
non-BPS D9-branes in type IIA string theory is given by\cite{Ho,KuMaMo2}
\begin{eqnarray}
T(x^\mu)&=&u\sum_{i=p+1}^9(x^i-\phi^i(x^\alpha))\,\gamma^i,\\
A_\alpha(x^\mu)&=&A_\alpha(x^\alpha)~~~(\alpha=0,\dots,p),~~~~
A_i(x^\mu)=0~~~(i=p+1,\dots,9).
\end{eqnarray}
Here $T(x^\mu)$ and $A_\mu(x^\mu)$ are the tachyon and gauge fields
on the non-BPS D9-brane, and $\phi^i(x^\alpha)$ and
$A_\alpha(x^\alpha)$ are the fluctuations corresponding to
scalar and gauge fields on the D$p$-brane.
Then, using (\ref{bpm})--(\ref{shift}),
we have
\begin{eqnarray}
e^{-S_b}\ket{B9;\pm}_{\rm NS}
&=&
\int[d\bx^\mu][d\bG^i]\,
\exp\left\{
\int d\wh\sigma\left(
\frac{1}{4}\bG^i D\bG^i
+u(\bx^i-\phi^i(\bx^\alpha))\bG^i
\right.\right.
\nn\\
&&\left.\left.\frac{}{}
\hspace{10ex}
-A_\alpha(\bx^\alpha)D\bx^\alpha-i\bx^\mu\bP_\mu
\right)\right\}\ket{\bx^\mu=0;\pm}_{\rm NS}\\
&\ra&\int[d\bx^\mu]\,\delta\left(\bx^i-\phi^i(\bx^\alpha)\right)\nn\\
&&~~~~\times
\exp\left\{-
\int d\wh\sigma\left(\,
A_\alpha(\bx^\alpha)D\bx^\alpha+i\bx^\mu\bP_\mu\,
\right)\right\}\ket{\bx^\mu=0;\pm}_{\rm NS}\nn\\
\label{delta}\\
&=&\exp\left\{-
\int d\wh\sigma\left(A_\alpha(\bX^\alpha)D\bX^\alpha
+i\phi^i(\bX^\alpha)\bP_i
\right)\right\}\ket{Bp;\pm}_{\rm NS}
\label{Dp}
\end{eqnarray}
This time, we performed $\bG^i$ integral and took $u\ra\infty$
limit in (\ref{delta}).
\footnote{A handy way to obtain
 (\ref{delta}) is to
rescale $u\bG^i\ra\bG^i$ and drop the kinetic term
of $\bG^i$ by taking the limit $u\ra\infty$ before
performing the $\bG^i$ integral.} \ 
In (\ref{Dp}), we again correctly reproduced
the D$p$-brane boundary state with the boundary interaction.

\para{Generalization to type I string theory}

Though we do not have enough space to explain in detail,
the whole story can also be elegantly
applied to type I string theory.
The gauge group and the representation of
tachyon and scalar fields are
much more complicated than those in type II
string theory as listed
in table \ref{table:1}.\cite{Be} \ 
In this table, $p=1,5,9$ are BPS D$p$-branes while
the others are non-BPS D$p$-branes.
\begin{table}[htb]
\vspace*{1ex}
\caption{D$p$-branes in type I string theory}
\vspace{1ex}
\label{table:1}
\begin{center}
\footnotesize
$
\begin{array}{c|ccccccccccc}
\hline\hline
p&-1&0&1&2&3&4&5&6&7&8&9\\
\hline
{\rm Gauge}& U&O&O&O&U&Sp&Sp&Sp&U&O&O\\
{\rm Tachyon}&\asym&\asym&--&\sym&\sym&\sym
&--&\asym&\asym&\asym&--\\
{\rm Scalar}&\mbox{adj.}&\sym&\sym&\sym&\mbox{adj.}
&\asym&\asym&\asym&\mbox{adj.}&\sym&--\\
\hline
\end{array}
$
\end{center}
\end{table}

As we can see from the table,  K-matrix theory based on
the D-instantons ($p=-1$ on the table) is similar
to that based on type IIA non-BPS D-instantons, but
now the tachyon is in the anti-symmetric representation.
We can construct D-branes from this matrix theory
in an analogous way as above.
Examining the fluctuations around the D-brane solutions,
we correctly recover all the ingredients listed
in the table \ref{table:1}. See Refs.~\citen{AsSuTe2,AsSuTe3}
for details.

\section{CS-term and index theorem}

So far, we have considered NS-NS sector boundary states.
But, our argument is also applicable to the R-R sector as well.
In particular, we can also precisely reproduce CS-term of
the D$p$-brane from the CS-term of K-matrix theory. 
The CS-term of K-matrix theory is given as
\begin{eqnarray}
S_{CS}^{D(-1)}(C,\Phi^\mu,T)=
\bra{C}e^{-S_b(\Phi^\mu,T)}\ket{B(-1);+}_{RR},
\label{KCS}
\end{eqnarray}
where $\bra{C}$ represents a closed string state corresponding
to the massless RR fields. Here the boundary interaction $e^{-S_b}$
in the R-R sector
is given by replacing the trace `$\Tr$' with supertrace
`$\Str$' in the previous expression (\ref{supertrP}).
If we insert the D$p$-brane configuration
(\ref{fluc2-1})--(\ref{fluc2-3}) in (\ref{KCS}),
using the analogous relation
as (\ref{Dp2}) for the R-R sector boundary states,
we obtain
\begin{eqnarray}
S_{CS}^{D(-1)}(C,\Phi^\mu,T)\ra
\bra{C}e^{-S_b(A_\mu,\phi^i)}\ket{Bp;+}_{RR}=
S_{CS}^{Dp}(C,A_\alpha,\phi^i)
\label{CSascent}
\end{eqnarray}
in the $u\ra\infty$ limit.

As an interesting corollary of this,
we naturally obtain a physical derivation of index theorem.

Here we consider type IIB string theory and take
 D-instanton - anti D-instanton system as the starting point.
The CS-term (\ref{KCS}) can be written more explicitly
as follows. First we introduce fermionic operators
$\psi^\mu_\pm\equiv\frac{1}{\sq}(\psi_0^\mu\pm i\wt\psi_0^\mu)$
satisfying
\begin{eqnarray}
\{\psi_\pm^\mu,\psi_\pm^\nu\}=0,~~~
\{\psi_\pm^\mu,\psi_\mp^\nu\}=\delta^{\mu\nu}.
\end{eqnarray}
and the Fock vacuum $\bra{-}$ and $\ket{+}$ annihilated by
$\psi_-^\mu$ and $\psi_+^\mu$, respectively,
with $\bracket{-}{+}=1$. Then, the CS-term is
\footnote{In this section, we ignore possible
$\alpha'$ corrections which do not
contribute to the topological invariants that we
are focusing on.}
\beq
S_{CS}^{D(-1)}=
\int d^{10}k\,\bra{-}
C(k_\mu,\psi_+^\mu)
 \Str \left(e^{-ik_\mu\wh\Phi^\mu+2\pi\wh\cF}\right)
\ket{+}, \label{CSDin}
\eeq
where
\begin{eqnarray}
C(x^\mu,\psi_+^\mu)&=&\sum_{n:{\rm even}}
C_{\mu_1\cdots\mu_n}(x^\mu)\,\psi_+^{\mu_1}\cdots\psi_+^{\mu_n}
=\int d^{10}k\,e^{-ik_\mu x^\mu} C(k_\mu,\psi_+^\mu)
\end{eqnarray}
is a formal sum over
the massless RR $n$-form fields $C_{\mu_1\cdots\mu_n}(x)$
and
\beq
\wh\Phi^\mu&=&\mat{\Phi^\mu,,,\wt\Phi^\mu},\\
\wh\cF&=&
\mat{-TT^\dag+\frac{1}{8\pi^2}
{[}\Phi^\mu{,}\Phi^\nu{]}\psi_-^\mu\psi_-^\nu
,-\frac{i}{2\pi}(\Phi^\mu T-T\wt\Phi^\mu)\psi_-^\mu
,-\frac{i}{2\pi}(\wt\Phi^{\mu}T^\dag-T^\dag \Phi^{\mu})\psi_-^\mu,
-T^\dag T+\frac{1}{8\pi^2}
{[}\wt\Phi^\mu{,}\wt\Phi^\nu{]}\psi_-^\mu\psi_-^\nu}.
\label{IIBF}
\eeq
This expression can be obtained by either
directly manipulating
(\ref{KCS}) or using T-duality
relations from the CS-term of the D9-\AD9 system given
in Refs.~\citen{KeWi,KrLa,TaTeUe}.

Let us assume that the RR-field is constant ($i.e.$ $k_\mu=0$)
and all the components except for the zero-form part $C_0$ are
zero. Then the CS-term takes very simple form,
since $\Phi^\mu$ do not contribute to it in this case.
Actually, it becomes the index of the tachyon operator,
\begin{eqnarray}
S_{CS}^{D(-1)}&=&C_0\Str\left(e^{-2\pi Q^2}\right)
=C_0\Tr\left(\sigma^3 e^{-2\pi Q^2}\right)\nn\\
&=&C_0
\left(\dim {\rm Ker}\, TT^\dag-\dim {\rm Ker}\, T^\dag T
\right)
=C_0\,{\rm Index}\,Q,
\label{indexQ}
\end{eqnarray}
where $Q \equiv \left(~~~T\atop T^\dag~~~\right)$.
Since the coupling of the RR $0$-form field represents
D-instanton charge, we conclude that the index of the tachyon
operator is interpreted as the D-instanton charge.
This can also be seen from the fact that
$\dim {\rm Ker}\, TT^\dag$ and $\dim {\rm Ker}\, T^\dag T$
correspond to the number of D-instantons
and anti D-instantons which are not annihilated by
the tachyon condensation, respectively.

Now consider the D-instanton charge in
the presence of BPS D$p$-branes  ($p={\rm odd}$).
The tachyon configuration representing
D$p$-branes in type IIB K-matrix theory is also
given as (\ref{fluc2-1}), though $\gamma^\alpha$
are no longer hermitian matrices,
and we have
\beq
Q=u \sum_{\alpha=0}^p
(\wh{p}_\alpha -iA_\alpha(\wh{x}) )\Gamma^{\alpha}
\equiv -iu \Dslash~,
\label{QDirac}
\eeq
where $\Gamma^{\alpha}=\left({~~~\gamma^\alpha
\atop\gamma^{\alpha\dag}~~~}\right)$ ~($\alpha=0,\dots,p$)
are $SO(p+1)$ gamma matrices.
Note that $\sigma^3$ in (\ref{indexQ}) can be regarded as
the chirality operator.
Therefore, from the above argument,
the D-instanton charge in the presence of
the BPS D$p$-branes is just the index of the usual
Dirac operator $\Dslash$ on the world-volume of the
D$p$-branes.

On the other hand, as we have seen in (\ref{CSascent}),
the CS-term of D-instanton - anti D-instanton system
with the D$p$-brane configuration is equal to
the CS-term of D$p$-brane.
In our case with $C=C_0={\rm constant}$,
the CS-term of D$p$-brane $S_{CS}^{Dp}$ is given as
\beq
S_{CS}^{Dp}=C_0 \int_{Dp} \tr e^{F/2\pi}.
\label{indexth}
\eeq
Comparing this with (\ref{indexQ}), we obtain
\begin{eqnarray}
{\rm Index}\,(-i\Dslash)=\int_{Dp}\tr e^{F/2\pi}.
\label{AtySin}
\end{eqnarray}
This is nothing but the Atiyah-Singer index
theorem.\cite{Atiyah-Singer} \ 
It is quite interesting that we have naturally reached
this result considering D-brane physics.
The physical interpretation is now clear.
The Dirac operator is the tachyon operator
which represent the D$p$-brane in the
D-instanton - anti D-instanton system
and its index gives the D-instanton charge.
The D-instanton can also be constructed as the instanton
configuration in the gauge theory on the D$p$-brane
world-volume and the instanton number is given by
the Chern number of the gauge bundle.
And the two descriptions actually agree
as expected.

\section{Conclusions}

In this paper, we have explained how to
describe D-branes in K-matrix theory.
We examined the fluctuations around the D$p$-brane
solution and showed that they correctly reproduce the fields
on the D$p$-brane. Plugging the D$p$-brane configurations
into the action of K-matrix theory, we precisely obtained
the D$p$-brane action as the effective action of
the fluctuations.
Our method can easily be applied to construction of D-branes in
a wide variety of unstable D-brane systems and provides
a simple derivation of D-brane ascent/descent relations.

In K-matrix theory, D-branes are described in terms
of the operators $T$ and $\Phi^\mu$ acting on a Hilbert space.
We translated this analytic description
to a geometric description of the D-branes
given in terms of a gauge theory on the world-volume manifold
of the D-branes. The equivalence of the analytic and geometric
descriptions of the D-branes follows from
the equivalence between path integral and
operator formulation of the boundary quantum mechanics.
As a corollary, the Atiyah-Singer index theorem is naturally
obtained by looking at the coupling to RR-fields.

It is now clear that any D-branes can be constructed as
a kind of bound states of unstable D-instantons.
Therefore, we can in principle study anything
about D-branes by studying K-matrix theory,
just as some particle physicists sometimes say
``the theory of elementary particles is supposed to be
a theory of everything, since everything in our world
is made by the elementary particles".
As we have shown in this paper,
we precisely recover D-brane action from K-matrix theory.
Thus, at least, any analysis based on the D-brane action can be
recovered in the context of K-matrix theory.

Note however that our consideration so far
 is limited within the classical (disk) level.
The next important step would be to consider the quantum
effects in K-matrix theory.

\section*{Acknowledgements}
We would like to thank the organizers of
the Nishinomiya-Yukawa
Memorial Symposium for giving us a chance to present
our work.
The work of S.S. is supported in part by
Danish Natural Science Research Council.
The work of T.A. is supported in part by
JSPS Research Fellowships for Young Scientists.


\begin{thebibliography}{99}

\bibitem{AsSuTe3}
T.~Asakawa, S.~Sugimoto and S.~Terashima,
``Exact Description of D-branes via Tachyon Condensation,''
JHEP {\bf 02} (2003), 011, hep-th/0212188.

\bibitem{BFSS}
T. Banks, W. Fischler, S.H. Shenker and L. Susskind,
``M Theory as a Matrix Model: A Conjecture,''
Phys. Rev. {\bf 55} (1997), 5112, hep-th/9610043.

\bibitem{IKKT}
N.~Ishibashi, H.~Kawai, Y.~Kitazawa and A.~Tsuchiya,
``A Large-N Reduced Model as Superstring,''
Nucl. Phys. {\bf B498} (1997), 467, hep-th/9612115.

\bibitem{AsSuTe}
T.~Asakawa, S.~Sugimoto and S.~Terashima,
``D-branes, matrix theory and K-homology,''
JHEP {\bf 03} (2002), 034, hep-th/0108085.

\bibitem{Connes}
A. Connes,
``Noncommutative geometry,'' Academic Press, 1994.

\bibitem{AsSuTe2}
T.~Asakawa, S.~Sugimoto and S.~Terashima,
``D-branes and KK-theory in type I string theory,''
JHEP {\bf 05} (2002), 007, hep-th/0202165.

\bibitem{WittenK}
E. Witten,
``D-branes and K-theory,''
JHEP {\bf 12} (1998), 019, hep-th/9810188.

\bibitem{Ho}
P. Horava,
``Type IIA D-Branes, K-Theory, and Matrix Theory,''
Adv. Theor. Math. Phys. {\bf 2} (1999), 1373, hep-th/9812135.

\bibitem{Te}
S. Terashima,
``A Construction of Commutative D-branes from Lower Dimensional
Non-BPS D-branes,''
JHEP {\bf 05} (2001), 059, hep-th/0101087.

\bibitem{KuMaMo2}
D. Kutasov, M. Marino and G. Moore,
``Remarks on tachyon condensation in superstring field theory,''
hep-th/0010108.

\bibitem{KrLa}
P. Kraus and F. Larsen,
``Boundary String Field Theory of the DDbar System,''
Phys. Rev. {\bf D63} (2001), 106004, hep-th/0012198.

\bibitem{TaTeUe}
T. Takayanagi, S. Terashima and T. Uesugi,
``Brane-Antibrane Action from Boundary String Field Theory,''
JHEP {\bf 03} (2001), 019, hep-th/0012210.

\bibitem{AnTs}
O.~D.~Andreev and A.~A.~Tseytlin,
``Partition Function Representation For The Open Superstring Effective Action:
Cancellation Of Mobius Infinities And Derivative Corrections To
Born-Infeld Lagrangian,''
Nucl.\ Phys.\ B {\bf 311} (1988) 205.

\bibitem{Wy}
N. Wyllard,
``Derivative corrections to D-brane actions
with constant background fields,''
Nucl. Phys. {\bf B598} (2001) 247,
hep-th/0008125.

\bibitem{CLNY}
C.~G.~Callan, C.~Lovelace, C.~R.~Nappi and S.~A.~Yost,
``Adding Holes And Crosscaps To The Superstring,''
Nucl. Phys. {\bf B293} (1987), 83;
``Loop Corrections to Superstring Equations of Motion,''
Nucl. Phys. {\bf B308} (1988), 221.

\bibitem{HaKuMa}
J.~A.~Harvey, D.~Kutasov and E.~J.~Martinec,
``On the relevance of tachyons,''
hep-th/0003101.

\bibitem{Be}
O. Bergman,
``Tachyon Condensation in Unstable Type I D-brane Systems,''
JHEP {\bf 11} (2000), 015, hep-th/0009252.

\bibitem{KeWi}
C.~Kennedy and A.~Wilkins,
``Ramond-Ramond couplings on brane-antibrane systems,''
Phys. Lett. {\bf B464} (1999), 206, hep-th/9905195.

\bibitem{Atiyah-Singer}
M. F. Atiyah and I. M. Singer,
``The index of elliptic operators I, III,''
Ann. of Math. {\bf 87} (1968), 484; 546.

\end{thebibliography}
\end{document}